%% file: main.tex
\documentclass[nofootinbib,showkeys,aps,prb,twocolumn,showpacs,twoside,letterpaper]{revtex4-1}
\input{local.tex}
\date{June 12, 2012}
\begin{document}
\title{Matrices, bases and matrix elements for cubic double crystallographic groups.}
\author{Oleg Chalaev}
\homepage[Receive announcements about my new articles:\\]{http://scholar.google.com/citations?user=9zdm3gEAAAAJ}
\email[E-mail me at: ]{chalaev@gmail.com}
\affiliation{Department of Physics and Astronomy, University of Missouri, Columbia, MO 65211}
\thanks{Helpful discussions with M.~E.~Flatté and G.~Vignale and the support of ARO MURI through Grant No. W911NF-08-1-0317 are gratefully acknowledged.}
\begin{abstract}
Matrices of the irreducible representations of double crystallographic point groups $O$, $T_d$, $O⊗\{𝟙,\hat I\}$ and $T_d⊗\{𝟙,\hat I\}$ are derived.
The characteristic polynomials (spinor bases) up to the sixth power are obtained.
The method for the derivation of the general form of an arbitrary matrix element of a vector/tensor quantity is developed;
as an application, the $\vec k·\hat{\vec p}$ matrix elements are calculated.
It is demonstrated that the other known  method for obtaining the bases of the irreducible representations of the double groups
($\hat{\vec L}\!·\!\hat{\vec S}$\,-diagonalization of a linear combination of spherical harmonics) is unreliable.
\end{abstract}
\maketitle

\section{Introduction}\label{sec:introduction}
The studies of the part of the group theory which is used in condensed matter physics
are commonly believed to be definitely finished by the middle of the XX century.
Introduction to group theory is now an indispensable part of condensed matter textbooks.
Surprisingly, some important results (e.g. the matrices for the double irreducible representations) are missing.
Consequently the power of group theory is not completely exploited.

Condensed matter textbooks often contain character tables (without matrices) of the double groups, and this might give a wrong impression that a character table provides
all symmetry group information one would ever need in condensed matter physics.
This might be true for a non-degenerate case: using character tables one easily understands whether a matrix element
$〈Ψ|\hat O|Φ〉$ of some operator $\hat O$ between two non-degenerate states $|Ψ〉$ and $|Φ〉$ must be zero by symmetry or not.
The situation becomes more complicated when the states $|Ψ〉$ and $|Φ〉$ are degenerate, and
$\hat O$ is a vector (or a higher order tensor).
In this case $〈Ψ|\hat O|Φ〉$ is a matrix (or a set of matrices), and one would like to understand its structure rather than just revealing if it is zero or not.
The knowledge of matrices allows one to get a deeper insight into $〈Ψ|\hat O|Φ〉$: one can deduce its most general form using straightforward
formalism, see Sec.~\ref{sec:SRmethod}.

Another usage of matrices is efficient constructing of the bases of irreducible representations (irreps).
It is possible\cite{PhysRevB.83.165210} to construct a \emph{polynomial} basis for an irreducible representation (irrep) without matrices:
one finds a linear combination of the spherical harmonics (which correspond to the same degenerate level of a lonely atom)
which diagonalizes the $\hat{\vec L}·\hat{\vec S}$ operator (where $\vec L$ is the angular momentum, and $\vec S$ denotes spin).
I see several drawbacks of this method though – unreliability, complexity, and narrowness:
\begin{itemize}
\item Unreliability: I am not aware of a rigid proof of reliability of this method (i.e., that every set of functions obtained using this method indeed does form a
basis of an irreducible representation). To me the requirement of the $\hat{\vec L}·\hat{\vec S}$ diagonalization seems too weak.
My suspects are confirmed in Sec.~\ref{sec:comparison} where an example of a wrong basis obtained with this method is presented.
\item Complexity: In order to obtain a high (say, sixth) power polynomial basis of an irrep one has to diagonalize a large matrix which may be hard to do
analytically.
\item Narrowness: the method is restricted to polynomial functions. These  can only \emph{approximate}  wave functions in the vicinity of zero, which is
unsatisfactory, e.g., in a numerical calculation where one would like to reconstruct the wave function (or electron density) in the entire atomic cell.
The knowledge of matrices permits construction of projection operators\cite{PeTri,DresselhausGT} which symmetry
\emph{arbitrary} (not necessary polynomial) wave functions. Such a symmetrization would allow, for example, to get rid of unphysical parts of the wave function
(which could appear, e.g., due to numerical errors), and might improve both accuracy and speed of numerical calculations.
\end{itemize}

Recently Elder et al.\cite{PhysRevB.83.165210} derived general form of the $\vec k·\hat{\vec p}$ Hamiltonian for the case of $T_d⊗\{𝟙,\hat I\}$  group (where $𝟙$
and $\hat I$ are unity and spatial
inversion operators, and $⊗$ stands for the direct product) using physical approach (diagonalizing the Hamiltonian in the presence of spin-orbit interaction).
In this article I solve the same problem using
projection operators approach\cite{PeTri,DresselhausGT} which seems me simpler, easier to check by the reader, and more reliable.

The calculation of matrices of  irreducible representations is based on the work of Dixon\cite{Dixon}, who is using the Burniside theorem,
according to which every irreducible representation (irrep) of a group is contained in some direct product of certain number of its \emph{faithful} (but not
necessary irreducible) representations. Dixon\cite{Dixon} demonstrated that in order to simultaneously block-diagonalize the set of direct products of matrices,
it is enough to diagonalize one (specially prepared) matrix.

The practical realization of the Dixon's method, however, may be problematic if the dimension of the direct product of representations is not small enough:
the problem arises due to the fact that it may be hard to diagonalize a large matrix analytically.
I faced this problem when dealing with double groups with inversion, $O⊗\{𝟙,\hat I\}$ and $T_d⊗\{𝟙,\hat I\}$, which have no faithful irreps in the
standard (Pauli) gauge  (which assumes that the inversion operator~$\hat I$ multiplies a spinor by $-1$).
In order to resolve it I had to use somewhat less known Cartan gauge, see Sec.~\ref{sec:inversion}.
This allowed me to obtain the matrices of the irreps analytically, but changed the double groups (including their character tables,
cf. Table~\ref{tab:TdCharactersCartan} and Ref.~[\onlinecite{YuCardona}]).

The choice of the gauge affects matrices of irreps and the character table but does not affect\cite{Altmann,Zhelnorowich,Joshi} physical quantities, e.g.,
bases of the irreps as well as matrix elements of operators (such as  $\vec k·\hat{\vec p}$) between the bands in a crystal.
(Note, however, that matrices of an irrep are always defined up to a unitary transformation; so
the corresponding basis sets are not unique.)

In order to save space I have included only two generators for each faithful irreducible representation in the text.
However, the reader is encouraged to use supplementary material\cite{suppMat}, where matrices for
all irreps  separated in classes together with the transformation parameters (see Sec.~\ref{sec:trPa}) are available.

It is important to check the obtained matrices. Fortunately, the check is much easier than the derivation; for convenience I provided a small  program which 
calculates characters and multiplication tables of the irreducible representations. 
The program can be easily expanded by the readers for additional tests.
It is available in the supplementary material\cite{suppMat}; there are versions for both {\tt maxima}\cite{maxima} and \emph{Mathematica}  computer algebra software (CAS) systems.

\section{Isomorphism between cubic ($O$) and tetrahedral ($T_d$) groups}\label{sec:inversion}
The simple (geometrical) group $O$ consists of all proper rotations\footnote{That is, transformations that conserve scalar product and do not alter the  basis
  signature.} of a cube and contains 24 elements divided in five classes as follows:
\newcolumntype{C}{>{$}c<{$}}
\begin{equation}\label{O:classes}    \begin{tabular}{|c|C|C|C|C|C|}\hline
class number →&1&2&3&4&5\tabularnewline\hline
\#\ of elements →&1&3&6&6&8\tabularnewline\hline
rotation axis →&
&\begin{minipage}{4.5ex}  \begin{center}    [110]\\ \insfig{4ex}{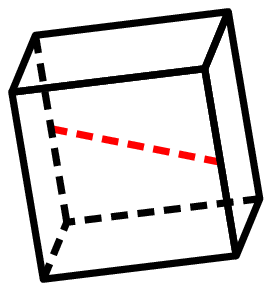}  \end{center}\end{minipage}
&\begin{minipage}{4.5ex}  \begin{center}    [100]\\ \insfig{4ex}{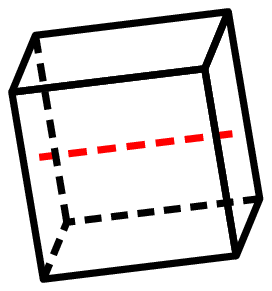}  \end{center}\end{minipage}
&\begin{minipage}{4.5ex}  \begin{center}    [100]\\ \insfig{4ex}{c3}  \end{center}\end{minipage}
&\begin{minipage}{4.5ex}  \begin{center}    [111]\\ \insfig{4ex}{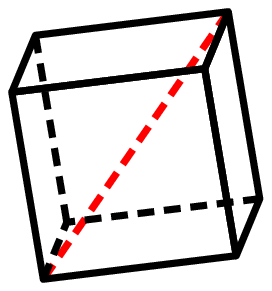}  \end{center}\end{minipage}\tabularnewline\hline
rotation angle →&0&π&π/2&π&2π/3\tabularnewline\hline
$Γ₅$ character→&3&-1&1&-1&0\tabularnewline\hline
\end{tabular}\end{equation}
The last line in the Table~\eqref{O:classes} is the character of the (faithful) $Γ₅$ irrep, which is given by the usual 3D proper rotation
matrices~\eqref{usualRotation} which transform a cube into itself. %

The full  tetrahedron symmetry group $T_d$ contains both proper and improper rotations
[the latter are emphasized by the overline and red color in Table~\eqref{Td:classes}].
The corresponding 3D rotation matrices compose (faithful) $Γ₄$ irrep.
The group $T_d$ is classified similarly to the $O$-group:
\begin{equation}\label{Td:classes}
    \begin{tabular}{|c|C|C|C|C|C|}\hline
class number →&1&2&3&4&5\tabularnewline\hline
\#\ of elements →&1&3&6&6&8\tabularnewline\hline
rotation axis →&
&\begin{minipage}{4.5ex}  \begin{center}    [001]\\ \insfig{4ex}{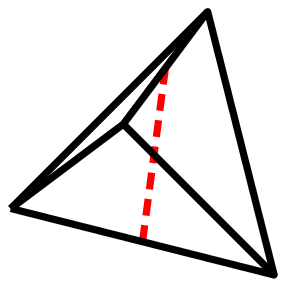}  \end{center}\end{minipage}
&\begin{minipage}{4.5ex}  \begin{center}    [001]\\ \insfig{4ex}{t2}  \end{center}\end{minipage}
&\begin{minipage}{4.5ex}  \begin{center}    [110]\\ \insfig{4ex}{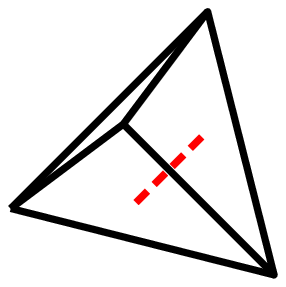}  \end{center}\end{minipage}
&\begin{minipage}{4.5ex}  \begin{center}    [111]\\ \insfig{4ex}{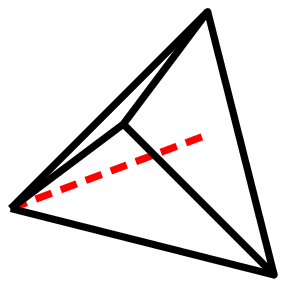}  \end{center}\end{minipage}\tabularnewline\hline
rotation angle\vphantom{\Big[}\ →&0&π&\color{red}\overline{π/2}&\color{red}\bar{π}&2π/3\tabularnewline\hline
$Γ₄$ character→&3& - 1& - 1& 1& 0\tabularnewline\hline
\end{tabular}\end{equation}
where the last line is the character of the $Γ₄$ irrep; differently from the $Γ₅$ irrep, $Γ₄$ contains matrices having determinant $-1$.
One can directly check that that the transformations of two \emph{simple} groups~$O$ and~$T_d$ obey the same multiplication table so that the two groups are
isomorphic.

The matrices of the other irreps ($Γ₁$,$Γ₂$, and $Γ₃$) are known\cite{PeTri}; otherwise one could deduce them
 from the direct products of the known \emph{faithful} irreps using the Dixon method\cite{Dixon}.

While the spatial inversion transformation operator $\hat I$ is uniquely defined in a 3D vector space ($\hat I=-𝟙$), the transformation of a spinor might be
different:\cite{Altmann,Zhelnorowich,Joshi}
$\hat I$ multiplies a spinor by a constant which may have any out of four values:  $±1$ (so called Pauli gauge) or  $±i$ (Cartan gauge).
For the groups without inversion center ($O$ and $T_d$), the choice $-1$ is especially convenient since it is the only one which leads to isomorphism
between the \emph{double} groups\footnote{The isomorphism between ``single'' (geometric)  groups does not necessary lead the isomorphism between the
  corresponding double groups.} $T_d$ and $O$.
On the other hand, Cartan gauge appears more  convenient for studying double groups $O⊗\{𝟙,\hat I\}$ and $T_d⊗\{𝟙,\hat I\}$ because it provides several
\emph{faithful} irreps to both groups. (In Pauli gauge these double groups posses no faithful irreps.)
While I would prefer to perform all calculations in the (standard) Pauli gauge,
the advantage of Cartan gauge is crucial: it strongly reduces the dimension of the matrices which I  had to diagonalize analytically while using the Dixon method\cite{Dixon}.
For this reason, both gauges are used in this article:
the results for double group $O$ are obtained in Pauli gauge;
the double group $T_d$ has been studied in both gauges: while matrices of the irreps and the character tables are gauge-dependent, the physical results
are\cite{Checked} gauge-invariant. Finally, I was unable\footnote{True Pauli-gauge-fans can nevertheless deduce the matrices in the Pauli gauge from my bases using
  the approach from Ref.~[\onlinecite{DresselhausGT}] if they really want.} to obtain analytical results for the double groups~$O⊗\{𝟙,\hat I\}$ 
and~$T_d⊗\{𝟙,\hat I\}$ in Pauli gauge; for these last two double groups Cartan gauge ($\hat I=i$ in spinor space) was used.

\section{Obtaining the matrices for the irreps}\label{sec:trPa}
An arbitrary element (transformation) of a double group can be characterized by the following (total five) parameters:
(i) three Euler angles~$(α,β,γ)$ which determine the rotation~\eqref{usualRotation} in the 3D space, (ii) the sign (or the branch number) in~\eqref{dude}, and
(iii) the presence/absence of inversion.

Below I provide a couple of generators for every faithful double irrep.
Complete sets of matrices grouped in classes can be found in the supplementary material\cite{suppMat}.%

\subsection{Double groups $O$ and $T_d$}\label{sec:allMatricesO}
In this section I use Pauli gauge so that the considered double groups are isomorphic and their matrices are the same (or similar).
The matrices for the first five irreps are the same as for the corresponding single (geometric) groups, see, e.g., Ref.~[\onlinecite{PeTri}].

There are several definitions of  Euler angles;
I use the following connection between Euler angles and 3D rotation matrix, see ([\onlinecite{Varsh}]1.4.54) or ([\onlinecite{Varsh}]1.4.63):
\begin{equation}\label{usualRotation}
    R_{\mathrm{3D}}^{α,β,γ}=R_{\mathrm{3D}}^z(α)R_{\mathrm{3D}}^y(β)R_{\mathrm{3D}}^z(γ),
\end{equation}
where
\begin{equation}\begin{split}
  R_{\mathrm{3D}}^z(α)=&\begin{pmatrix}\cosα&-\sinα&0\cr\sinα&\cosα&0\cr0&0&1\end{pmatrix},\quad\text{and}\\
  R_{\mathrm{3D}}^y(β)=&\begin{pmatrix}\cosβ&0&\sinβ\cr0&1&0\cr-\sinβ&0&\cosβ&\end{pmatrix}.
\end{split}  \end{equation}
It is straightforward to obtain
Euler angles for all proper rotations which transform a cube into itself. Substituted in~\eqref{usualRotation}, these Euler angles would produce
24 matrices of the $Γ₅$-representation.
Similarly, the set of matrices for the $Γ₄$-irrep can be obtained from all rotations which transform a tetrahedron into itself.
[If the rotation is improper, the corresponding 3D rotation matrix in~\eqref{usualRotation} changes its sign.]

Matrices for the $Γ₆$ representation can be obtained by substituting the values of $α,β,γ$ for the $O$-group into
the expression  ([\onlinecite{Varsh}]2.5.32) for the spinor rotation operator:
\begin{equation}\label{dude}
  D^{1/2}(α,β,γ)=
∓  \begin{pmatrix}
    \cos\frac{β}2\,e^{-i\frac{α+γ}2}&\sin\frac{β}2\,e^{-i\frac{α-γ}2}\cr
   -\sin\frac{β}2\,e^{i\frac{α-γ}2} &\cos\frac{β}2\,e^{i\frac{α+γ}2}
  \end{pmatrix},\end{equation}
where I have inserted $∓$ which stands for two branches of the matrix function~$D^{1/2}$.
[In physics textbooks a double group is often defined with the concept of ``non-identical rotation by $2π$'';
I prefer the more formal ``two branches'' definition\cite{Opechowski} instead.]
Note that in Pauli gauge used in this section the rotation parameters of the $T_d$ group generate the same set of matrices, as the parameters for the $O$-group.

I had to insert $∓$ instead of $±$ in~\eqref{dude} in order to achieve the compatibility with Ref.~[\onlinecite{YuCardona}], where the class $\{6S₄\}$
in the character table of the double $T_d$-group is understood to be composed of the \emph{first}-branch matrices.

Both branches of~\eqref{dude} produce 48 different $2×2$ matrices which compose the so-called $Γ₆$ representation.
Using CAS one easily separates these matrices into eight classes and obtains the multiplication table which demonstrates that $Γ₆$ is a faithful irrep.

All irreps of a finite group are contained\cite{Dixon} in a certain direct product of its faithful representations;
this means that it is enough to know only one faithful  (but not necessary irreducible) representation of a finite group in order to derive (at least numerically) matrices
for  all its irreducible representations.
The search of irreps is further simplified if we know the character table\cite{YuCardona} for the double group, which tells us that
all the missing (non-trivial) irreps (that is, $Γ₂,Γ₃,Γ₇,Γ₈$) are contained in the following direct products:
\begin{equation}
    Γ₄⊗Γ₆=Γ₇+Γ₈,\quad Γ₄⊗Γ₅=Γ₂+Γ₃+Γ₄+Γ₅.
\end{equation}
The extraction algorithm (that is, the simultaneous transformation of all matrices of a reducible representation into block diagonal structure) is invented and explained by Dixon\cite{Dixon}.

Since the double groups $O$ and $T_d$ are isomorphic their irreps-matrices can be chosen to be the same.
However, these two double groups are not identical (e.g., their irreps have different bases).
The reason for this discrepancy is that coordinate functions are transformed differently in these groups
(for $O$, the corresponding irrep is $Γ₅$, while for $T_d$ it is $Γ₄$).
Due to the same reason the double groups
$O⊗\{𝟙,\hat I\}$ and $T_d⊗\{𝟙,\hat I\}$ also have different bases, no matter what gauge (Pauli or Cartan) is used for the spatial inversion operator~$\hat I$.

The generators of the faithful irreps $Γ₆$, $Γ₇$, and $Γ₈$ are:
\begin{itemize}
\item for $Γ₆$:
\begin{equation}
    \frac1{√2} \begin{pmatrix}1&-1\\1&1 \end{pmatrix}\quad\text{and}\quad\frac1{√2}\begin{pmatrix}√i&√i\\-√{-i}&√{-i} \end{pmatrix},
\end{equation}
where $√{±i}≡\exp[±iπ/4]$,
\item for $Γ₇$:
\begin{equation}
    \frac1{√2}\begin{pmatrix}-1&1\\-1&-1 \end{pmatrix}\quad\text{and}\quad\frac1{√2}\begin{pmatrix}√{-i}&√{-i}\\ -√i&√i \end{pmatrix},
  \end{equation}
\item for $Γ₈$:
\begin{equation}
  \begin{split}
 \frac1{2√2}\begin{pmatrix}1&-√3&√3&-1\\ √3&-1&-1&√3\\ √3&1&-1&-√3\\ 1&√3&√3&1\end{pmatrix}\quad\text{and}\\
 \frac1{2√2}\begin{pmatrix}-√{-i}&-√{-3i}&-√{-3i}&-√{-i}\\ -√{3i}&-√i&√i&√{3i}\\ √{-3i}&-√{-i}&-√{-i}&√{-3i}\\ √i&-√{3i}&√{3i}&-√i \end{pmatrix},
  \end{split}
  \end{equation}
\end{itemize}
where every pair of matrices has the following transformation parameters:
\begin{enumerate}
\item  $α=0,β=π/2,γ=0$, second (positive) branch in~\eqref{dude}, and
\item  $α=β=π/2,γ=π$, first (negative) branch in~\eqref{dude}.
\end{enumerate}

\subsection{Double groups $O⊗\{𝟙,\hat I\}$ and $T_d⊗\{𝟙,\hat I\}$}\label{sec:allMatricesOh}
All results of this section are obtained in Cartan gauge, see Sec.~\ref{sec:inversion}.
Both double groups, $O⊗\{𝟙,\hat I\}$ and $T_d⊗\{𝟙,\hat I\}$ have 16 classes (and the same number of irreps).
The groups are \emph{not} isomorphic – they have different character tables (see Tab.~\ref{tab:OhOCartan} and Tab.~\ref{tab:OhTdCartan}) and non-similar matrices.
The matrices for ten ``single'' irreps $Γ₁^±…Γ₅^±$ can be\cite{Checked} easily derived from the corresponding irreps $Γ₁…Γ₅$ (the ones for the groups $O$
and~$T_d$) as follows. For even representations\footnote{As we see below, the characteristic polynomials of irreps $Γ₆^±…Γ₈^±$ are even/odd with respect to the
  transformation $(x,y,z)→(-x,-y,-z)$, and the same is valid also for the irreps $Γ₁^±…Γ₅^±$.} $Γ₁⁺…Γ₅⁺$, both first  eight and last eight  classes are given by
the same matrices as for the group $O$, see Sec.~\ref{sec:allMatricesO}. The same is valid for the ``odd'' representations $Γ₁⁻…Γ₅⁻$, except for that the
matrices for the last eight classes are multiplied by $-𝟙$. These ``single'' irreps are identical for both double groups (in particular, the corresponding
matrices are the same (or similar) for $O⊗\{𝟙,\hat I\}$ and $T_d⊗\{𝟙,\hat I\}$).

Without knowing \emph{apriori} the character table generating the irreps-matrices is somewhat more complicated.
In case of $O⊗\{𝟙,\hat I\}$ we depart from the transformation parameters of the double group~$O$;
substituted in Eqs.~\eqref{usualRotation} and~\eqref{dude}, these parameters produce 48 (out of total 96) matrices of the irreps~$Γ₅⁻$ and~$Γ₆⁺$.
In case of $Γ₅⁻$, the rest of the matrices is obtained by multiplying the first 48 ones by~$-1$; in case of $Γ₆⁺$, the multiplication constant is~$-i$.

Recursively going over different direct products of the known irreps we inevitably obtain the matrices\cite{suppMat} for all 16 irreps.
The traces of the matrices composing the first 8 classes coincide with one of the lines in Table~\ref{tab:TdCharactersCartan};
the coinciding line determines the number (denoted by the subscript) of the newly obtained irrep.
In addition to the number of the irrep, we have to determine the parity (denoted by the $±$ superscript); I do this by building
polynomial bases and checking their parity under the transformation $(x,y,z)→(-x,-y,-z)$.

The generators of the six faithful irreps ($Γ₆^±$, $Γ₇^±$, and $Γ₈^±$) of the double group $O⊗\{𝟙,\hat I\}$ in Cartan gauge are:
\begin{itemize}
\item for $Γ₆⁺$:
\begin{equation}
    \frac1{√2} \begin{pmatrix}1&-1\\1&1 \end{pmatrix}\quad\text{and}\quad\frac1{√2}\begin{pmatrix}-√{-i}&-√{-i}\\-√i&√i \end{pmatrix},
  \end{equation}
\item for $Γ₆⁻$:
\begin{equation}
    \frac1{√2} \begin{pmatrix}1&1\\-1&1 \end{pmatrix}\quad\text{and}\quad\frac1{√2}\begin{pmatrix}-√i&√i\\-√{-i}&-√{-i} \end{pmatrix},
  \end{equation}
\item for $Γ₇⁺$:
\begin{equation}
    \frac1{√2}\begin{pmatrix}-1&-1\\1&-1 \end{pmatrix}\quad\text{and}\quad\frac1{√2}\begin{pmatrix}-√{-i}&√{-i}\\√i&√i \end{pmatrix},
  \end{equation}
\item for $Γ₇⁻$:
\begin{equation}
    \frac1{√2}\begin{pmatrix}-1&-1\\1&-1 \end{pmatrix}\quad\text{and}\quad\frac1{√2}\begin{pmatrix}-√i&√i\\√{-i}&√{-i} \end{pmatrix},
  \end{equation}
\item for $Γ₈⁺$:
\begin{equation}
  \begin{split}
    \frac1{√2} \begin{pmatrix}1&1&0&0\\-1&1&0&0\\0&0&-1&1\\0&0&-1&-1\end{pmatrix}\quad\text{and}\\
    \frac1{2√2}\begin{pmatrix}
      √{-i}&-√{-i}&√{3i}&-√{3i}\\
      -√i&-√i&-√{-3i}&-√{-3i}\\
      -√{-3i}&-√{-3i}&-√i&-√i\\
      -√{3i}&√{3i}&-√{-i}&√{-i}
    \end{pmatrix},
  \end{split}
  \end{equation}
\item for $Γ₈⁻$:
\begin{equation}
  \begin{split}
   \frac1{2√2} \begin{pmatrix}-2&0&1&-√3\\0&2&i√3&i\\-1&i√3&1&√3\\√3&i&√3&-1\end{pmatrix}\quad\text{and}\\
   \frac1{2√2}\begin{pmatrix}
      √i&√{-3i}&-√i&-√{3i}\\
      √{-3i}&√i&√{-3i}&-√{-i}\\
     2√{-i}&0&-√{-i}&√{-3i}\\
      0&2√i&-√{-3i}&-√{-i}
    \end{pmatrix},
  \end{split}
  \end{equation}
\end{itemize}
where every pair of matrices has the following transformation parameters:
\begin{enumerate}
\item $α=0,β=π/2,γ=0$, inversion is absent, second (positive) branch in~\eqref{dude}, and
\item $α=β=π/2,γ=π$, inversion is present, first (negative) branch in~\eqref{dude}.
\end{enumerate}
The corresponding characters are given in Table~\ref{tab:OhOCartan}.

The same procedure is used for calculating the irreps-matrices for the double group $T_d⊗\{𝟙,\hat I\}$, and the characters are given in Table~\ref{tab:OhTdCartan}.
The generators of the six faithful irreps ($Γ₆^±$, $Γ₇^±$, and $Γ₈^±$) of the double group $T_d⊗\{𝟙,\hat I\}$ are:
\begin{itemize}
\item for $Γ₆⁺$:
\begin{equation}
  \frac i{√2} \begin{pmatrix}-1&1\\ -1&-1 \end{pmatrix}\quad\text{and}\quad\frac1{√2}\begin{pmatrix}-√{-i}&-√{-i}\\ -√i&√i \end{pmatrix},
  \end{equation}
\item for $Γ₆⁻$:
\begin{equation}
   \frac i{√2} \begin{pmatrix}-1&-1\\1&-1 \end{pmatrix}\quad\text{and}\quad\frac1{√2}\begin{pmatrix}-√i&√i\\√{-i}&√{-i} \end{pmatrix},
  \end{equation}
\item for $Γ₇⁺$:
\begin{equation}
   \frac i{√2}\begin{pmatrix}1&1\\-1&1\end{pmatrix}\quad\text{and}\quad\frac1{√2}\begin{pmatrix}-√{-i}&√{-i}\\√i&√i \end{pmatrix},
  \end{equation}
\item for $Γ₇⁻$:
\begin{equation}
  \frac i{√2}\begin{pmatrix}1&1\\-1&1 \end{pmatrix}\quad\text{and}\quad\frac1{√2}\begin{pmatrix}-√i&√i\\√{-i}&√{-i} \end{pmatrix},
  \end{equation}
\item for $Γ₈⁺$:
\begin{equation}
  \begin{split}
   \frac i{√2} \begin{pmatrix}-1&-1&0&0\\1&-1&0&0\\0&0&1&-1\\0&0&1&1\end{pmatrix}\quad\text{and}\\
    \frac1{2√2}\begin{pmatrix}
      √{-i}&-√{-i}&√{3i}&-√{3i}\\
      -√i&-√i&-√{-3i}&-√{-3i}\\
      -√{-3i}&-√{-3i}&-√i&-√i\\
      -√{3i}&√{3i}&-√{-i}&√{-i}
    \end{pmatrix},
  \end{split}
  \end{equation}
\item for $Γ₈⁻$:
  \begin{equation}
    \label{eq:2}
    \begin{split}\frac1{2√2}\begin{pmatrix}-2i&0&i&-i√3\cr 0&2i&-√3&-1\cr -i&-√3&i&i√3\cr i√3&-1&i√3&-i  \end{pmatrix}
\text{and}\\
\frac1{2√2}\begin{pmatrix}√i&√{-3i}&-{√i}&-√{3i}\cr  √{-3i}&{√i}&√{-3i}&-√{-i}\cr 2√{-i}&0& -√{-i}& √{-3i}\cr  0&2√i& -√{-3i} & -√{-i}  \end{pmatrix},
 \end{split}  \end{equation}
\end{itemize}
where every pair of matrices has the following transformation parameters:
\begin{enumerate}
\item $α=0,β=π/2,γ=0$, inversion is absent, second (positive) branch in~\eqref{dude}, and
\item $α=β=π/2,γ=π$, inversion is present, first (negative) branch in~\eqref{dude}.
\end{enumerate}

\section{Generalized selection rules}\label{sec:GSR}
\subsection{The calculation method}\label{sec:SRmethod}
Let us study the system which symmetry is given by some double group~$G$ which contains~$|G|$ elements.
It is convenient to present matrix elements $〈Ψ|\hat O|Φ〉$ between degenerate bands in the matrix form.
[For example, a matrix element between a (two-fold) $Γ₆$-band and a (four-fold) $Γ₈$-band is a 2×4 matrix.]
Symmetry enforces restrictions on matrix elements $〈Ψ|\hat O|Φ〉$ of an operator $\hat O$ between the bands $|Ψ〉$ and $|Φ〉$.
Suppose that the states in the band $Ψ$ are transformed according to some irrep named~$A$, the operator $\hat O$ is transformed according to some irrep~$B$,
and the states in the band $Ψ$ are transformed according to some irrep~$C$.

The case when $\hat O$ is a scalar is trivial and will not be considered.
In case when $\hat O$ is a vector\footnote{The generalization for the case when~$\hat O$ is a tensor is straightforward.} operator (e.g., $\hat O≡\hat{\vec O}=\hat{\vec p}$\,) the quantity $〈Ψ|\hat O|Φ〉$ is characterized by three indexes
\begin{equation}\label{theIntegrand}
〈ijk〉≡〈Ψ_i|{\hat O}_j|Φ_k〉=∫Ψ^†_i(λ){\hat O}_jΦ_k(λ)\udλ,
\end{equation}
where~$λ$ represents all arguments of a wave function (except for spin) in some representation (e.g.~$λ$ may be coordinate or momentum).
In total~$〈Ψ|\hat O|Φ〉$ has $n=\dim A·\dim B·\dim C$ elements. The symmetry (i) requires that some of the matrix elements in~\eqref{theIntegrand} are zero, while
among the others only few (sometimes -- only one) are independent -- see, e.g.,  Eq.~\eqref{m647} below.

Under the action of a symmetry element $\hat g∈G$ the integrand in~\eqref{theIntegrand} is transformed\footnote{Following Refs.~\onlinecite{PeTri,Anselm}, I use the ``transposed''
  definition for the irreps matrices: $gΨ_i=∑_{i'}(A_g)_{i'i}Ψ_{i'}=(A_g^TΨ)_i$.} according to the direct product of three representations~$A⊗B⊗C$:
\begin{equation}\label{actG}
  \hat g〈ijk〉=∑_{i',j',k'}A^*_{i'i}(g)B_{j'j}(g)C_{k'k}(g)\,〈i'j'k'〉.
\end{equation}
The group averaging operator
\begin{equation}\label{gao}
{\hat{ℙ}}_G=|G|^{-1}∑_{\hat g∈G}\hat g
\end{equation}
commutes with the integration in~\eqref{theIntegrand} and leaves all matrix elements intact:
\begin{equation}
{\hat{ℙ}}_G〈ijk〉=〈ijk〉.
\end{equation}
Let us associate every set of indexes $(i,j,k)$ with some (orthonormal) bases element~$e_l$ of the
$n$-dimensional complex space~$ℂ^n$:
\begin{equation}\label{otoc}\begin{split}
    e_l↔(i,j,k)→〈ijk〉,\quad   l(i,j,k)=\\ =(i-1)·\dim B·\dim C+(j-1)·\dim C+k,
\end{split}\end{equation}
where we have taken into account the fact that different index sets $(i,j,k)$ may correspond to the
same values of matrix elements~$〈ijk〉$.

The transformation~\eqref{actG} corresponds to a linear operator in~$ℂ^n$:
\begin{equation}  \begin{split}
     \hat g(ijk)=∑_{i',j',k'}A^*_{i'i}(g)B_{j'j}(g)C_{k'k}(g)·(i'j'k')\\
     \text{or } \hat ge_l=∑_{l'}〈l'|\hat g|l〉e_{l'},
\end{split}\end{equation}
where we used one-to-one correspondence between $l$ and~$(i,j,k)$ defined in~\eqref{otoc}.
Similarly, the group averaging operator from~\eqref{gao} can be associated with a linear operator in~$ℂ^n$:
\begin{equation}\label{GinCn}
{\hat{ℙ}}_Ge_l=∑_{l'}〈l'|{\hat{ℙ}}_G|l〉e_{l'},\quad 〈l'|{\hat{ℙ}}_G|l〉=\frac1{|G|}∑_{g∈G}〈l'|\hat g|l〉.
\end{equation}
Then we consider ${\hat{ℙ}}_G$ as a linear operator which acts in the space $ℂ^n$:
\begin{equation}\label{mpinCn}
∀\vec v=∑_{i=1}ⁿa_ie_i∈ℂ^n\quad {\hat{ℙ}}_G\vec v∈ℂ^n.
\end{equation}
A direct product of irreps is often reducible. Reducibility means that there exists a unitary transformation which converts
all ($∀g∈G$) matrices $〈l'|\hat g|l〉$ from~\eqref{GinCn} into the block
diagonal form, and every block would correspond to some irrep of the group G.
The group theory tells us that the averaging $|G|^{-1}∑_{g∈G}〈l'|\hat g|l〉$ destroys (averages to zero) all matrix blocks except for
those (one-dimensional) ones, which correspond to the trivial irrep $Γ₁$ (or~$Γ₁⁺$).
In other words, $n$-dimensional matrix~${\hat{ℙ}}_G$ in~\eqref{mpinCn}  is similar to a diagonal matrix where the only non-zero elements are ones.

This means that the operator ${\hat{ℙ}}_G:ℝ^n→ℝ^m$ is a projector, and that it  has only two
different eigenvalues: 0 and 1. The degeneracy $m$ of the eigenvalue~1 is the same as the number of times which the trivial irrep $Γ₁$ (or~$Γ₁⁺$) enters in the direct product
\begin{equation}
  A⊗B⊗C=mΓ₁+\text{other irreps.}
\end{equation}
In case when $m=1$, all matrix elements $〈ijk〉$ are proportional to \emph{only one} constant.
The proportionality coefficients are just components of the eigenvector of the projector ${\hat{ℙ}}_G$ which corresponds to the eigenvalue 1.
In order to obtain the eigenvector, we choose any set~$(i',j',k')$ which is not projected to zero by ${\hat P}_G$.
In other words
\begin{equation}\label{razMat}
  〈ijk〉∝\frac1{|G|}∑_{g∈G}A^*_{i'i}(g)B_{j'j}(g)C_{k'k}(g)
\end{equation}
where I can take any set of indexes $(i',j',k')$ which produces a non-zero value in~\eqref{razMat}.
[In case when $m=0$, any set $(i',j',k')$ produces zero in~\eqref{razMat}.]
An example of the case $m=1$ is the matrix element of $\vec k·\hat{\vec p}$ between the $Γ₆$ conduction band and the $Γ₇$ valence band in GaAs.
Since the symmetry of GaAs is described by the double group $T_d$, the vector operator $\hat{\vec p}$ is transformed according to the irrep~$Γ₄$. From the fact that
\begin{equation}\label{oneConst}
  Γ₆⊗Γ₄⊗Γ₇=Γ₁+Γ₃+Γ₄+2Γ₅
\end{equation}
we conclude that~$m=1$.
One of the indexes set which produces a non-zero value in~\eqref{razMat} is $(i',j',k')=(1,1,1)$.
Substituting this and two other index sets into~\eqref{razMat} we obtain
\begin{equation}\label{explainME}  \begin{split}
&\text{for }(i',j',k')=(1,1,1)\quad    〈ijk〉∝i(\vec k·\vec{σ})σ₂/6,\\
&\text{for }(i',j',k')=(1,1,2)\quad    〈ijk〉∝0,\\
&\text{for }(i',j',k')=(1,2,1)\quad    〈ijk〉∝(\vec k·\vec{σ})σ₂/6,
  \end{split}\end{equation}
where~$\vec{σ}≡(σ₂,σ₂,σ₃)$ is the set of Pauli matrices.
Going over all possible values of $(ijk)$ we conclude that in a zinc-blende structure the matrix element
$〈Γ₆|\vec k·\hat{\vec p}\,|Γ₇〉$ is parametrized by \emph{one} constant, as predicted in~\eqref{oneConst}:
\begin{equation}\label{m647}
〈Γ₆|\vec k·\hat{\vec p}\,|Γ₇〉∝(\vec k·\vec{σ})σ₂.
\end{equation}
One may note that according to the bases written in Ref.~[\onlinecite{WinklerBasis}],
$〈Γ₆|\vec k·\hat{\vec p}\,|Γ₇〉∝\vec k·\vec{σ}$ and ask why there is an extra~$σ₂$ in~\eqref{explainME} and~\eqref{m647}.
This happens because my $Γ₇$ basis~\eqref{TdGammaVII} is different from the one in Winkler's book\cite{WinklerBasis}:
in order to obtain Winkler's basis one has to (i) exchange basis functions in~\eqref{TdGammaVII} and (ii) change the sign in front of one of the basis functions.
Since both of these operations correspond to unitary transformations of the irreps,
both  basis~\eqref{TdGammaVII} and the matrix~\eqref{m647} are in agreement with the bases in Ref.~[\onlinecite{WinklerBasis}].

In case when $m>1$,  matrix elements $〈ijk〉$ are parametrized by $m$ independent complex constants.
I illustrate this case on the example of $〈Γ₈|\vec k·\hat{\vec p}\,|Γ₈〉$. From the fact that
\begin{equation}
  Γ₈⊗Γ₄⊗Γ₈=2Γ₁+2Γ₂+4Γ₃+6Γ₄+6Γ₅
\end{equation}
we conclude that $m=2$ in this case, so that
\begin{equation}\label{m848}
  〈Γ₈|\vec k·\hat{\vec p}\,|Γ₈〉=c₁M₁+c₂M₂,\quad c₁,c₂∈ℂ.
\end{equation}
The generalized selection rules allow us to determine the matrices~$M₁$ and~$M₂$
following the same prescription which we used above in order to obtain the result~\eqref{m647} for $〈Γ₆|\vec k·\hat{\vec p}\,|Γ₇〉$.
Going over all possible values of $(ijk)$ we obtain several (more than $m=2$) \emph{linearly dependent}\footnote{I understand linear (in)dependence of matrices
as linear (in)dependence of corresponding vectors (matrix elements can always be rearranged in a vector).} non-zero matrices.
Since we know that $m=2$, it is enough to consider any two linearly independent matrices, e.g.,
For $(i',j',k')=(1,1,2)$ we obtain
\begin{align}\nonumber M₁∝
k₁\begin{pmatrix}0&5&0&√3\cr -3&0&-√3&0\cr 0&-√3&0&-3\cr √3&0&5&0\end{pmatrix}+\\
+k₂\begin{pmatrix}0&5\,i&0&-√3\,i\cr 3\,i&0&-√3\,i&0\cr 0&√3\,i&0&-3\,i\cr √3\,i&0&-5\,i&0\end{pmatrix}+\\
+k₃\begin{pmatrix}0&0&2&0\cr 0&0&0&6\cr -6&0&0&0\cr 0&-2&0&0\end{pmatrix}.
\nonumber\end{align}
For $(i',j',k')=(2,1,1)$ we obtain
\begin{align}\nonumber M₂∝
k₁\begin{pmatrix}0&-3&0&√3\cr 5&0&-√3&0\cr 0&-√3&0&5\cr √3&0&-3&0\end{pmatrix}+\\
+k₂\begin{pmatrix}0&-3\,i&0&-√3\,i\cr -5\,i&0&-√3\,i&0\cr 0&√3\,i&0&5\,i\cr √3\,i&0&3\,i&0\end{pmatrix}+\\
+k₃\begin{pmatrix}0&0&-6&0\cr 0&0&0&-2\cr 2&0&0&0\cr 0&6&0&0\end{pmatrix}.
\nonumber\end{align}
Any other set of values~$(i',j',k')$ produces\cite{Checked} a matrix which (together with $M₁$ and $M₂$) forms a linearly dependent system of matrices.
E.g., for $(i',j',k')=(3,2,2)$ we obtain
\begin{align}\nonumber M₃∝
k₁\begin{pmatrix}0&-√3\,i&0&-3\,i\cr -√3\,i&0&3\,i&0\cr 0&3\,i&0&-√3\,i\cr -3\,i&0&-√3\,i&0\end{pmatrix}+\\
+k₂\begin{pmatrix}0&√3&0&-3\cr -√3&0&-3&0\cr 0&3&0&√3\cr 3&0&-√3&0\end{pmatrix}+\\
+k₃\begin{pmatrix}0&0&2\,√3\,i&0\cr 0&0&0&-2\,√3\,i\cr 2\,√3\,i&0&0&0\cr 0&-2\,√3\,i&0&0\end{pmatrix}.
\nonumber\end{align}
Any two matrix sets from  $\{M₁,M₂,M₃\}$ are linearly independent, but the whole set $\{M₁,M₂,M₃\}$ is linearly dependent.
This means, e.g., that the following expression is equivalent to~\eqref{m848}:
\begin{equation}\label{altChoice}
  〈Γ₈|\vec k·\hat{\vec p}\,|Γ₈〉=c₁'M₁+c₂'M₃,\quad c₁',c₂'∈ℂ.
\end{equation}
In case when the matrix element $〈Γ₈|\vec k·\hat{\vec p}\,|Γ₈〉$ is calculated between \emph{the same}  $Γ₈$-bands, the
two \emph{complex}  constants in~\eqref{m848} or in~\eqref{altChoice} gain a constraint: their values  should be chosen in such a way that
$〈Γ₈|\vec k·\hat{\vec p}\,|Γ₈〉$ is Hermitian.
\vspace{2ex}\hrule\vspace{2ex}
In sections~\ref{s:a:kp:O}, \ref{s:a:kp:Td}, \ref{sec:OxI}, and~\ref{sec:TdxI} I present matrix elements of  the $\vec k·\hat{\vec p}$  operator which are
obtained using CAS according to the method described in Sec.~\ref{sec:SRmethod}.

\begin{table}  \centering    \begin{tabular}{|c|C|C|C|C|C|C|C|C|}\hline
class →&a&b&c&d&e&f&g&h\tabularnewline\hline
\#\ of elements →&1&6&6&12&8&1&6&8\tabularnewline\hline
rotation angle\vphantom{\Big[}\ →&0&\color{red}\overline{π}&\color{red}\overline{π/2}&π&2π/3&0&\color{red}\overline{π/2}&2π/3\tabularnewline\hline
rotation axis →&&\insfig{4ex}{t3} &\insfig{4ex}{t2}&\insfig{4ex}{t2}&\insfig{4ex}{t1}&&\insfig{4ex}{t2}&\insfig{4ex}{t1}\tabularnewline\hline
Γ₁&1&1&1&1&1&1&1&1\tabularnewline\hline
Γ₂&1&1&-1&-1&1&1&-1&1\tabularnewline\hline
Γ₃&2&2&0&0&-1&2&0&-1\tabularnewline\hline
Γ₄&3&-1&-1&1&0&3&-1&0\tabularnewline\hline
Γ₅&3&-1&1&-1&0&3&1&0\tabularnewline\hline
Γ₆&2&0&-i√2&0&1&-2&i√2&-1\tabularnewline\hline
Γ₇&2&0&i√2&0&1&-2&-i√2&-1\tabularnewline\hline
Γ₈&4&0&0&0&-1&-4&0&1\tabularnewline\hline
\end{tabular}\caption{Character table for the double group $T_d$ in Cartan gauge. The classes are enumerated in the same order as in
Ref.~[\onlinecite{YuCardona}]. The characters for $Γ₁…Γ₅$ and $Γ₈$ are the same as in Ref.~[\onlinecite{YuCardona}].
The characters for $Γ₆$ and $Γ₇$ differ from the standard\cite{YuCardona} ones by an extra factor $-i$ which appears in front of $√2$.
The 3D vectors are transformed according to the irrep $Γ₄$; for the spinors the appropriate irrep is $Γ₆$.
\label{tab:TdCharactersCartan}}
\end{table}

\subsection{Comparison with Elder et al.\cite{PhysRevB.83.165210}}\label{sec:comparison}
Elder et al.\cite{PhysRevB.83.165210} recently found $\vec k·\hat{\vec p}$ matrix elements using Löwdin approach which allows
to split (approximately) the Hamiltonian into several blocks in such a way, that each block corresponds to a separate energy level (or to an irrep).

Let us compare the result~\eqref{m848} with~([\onlinecite{PhysRevB.83.165210}]31).
We have to take into account the difference between the bases on p.~[\onlinecite{PhysRevB.83.165210}]19 and~\eqref{TdGammaVIII}.
One notices that the \emph{linear} basis from~\eqref{TdGammaVIII} is equivalent but not equal to the linear basis for the irrep $Γ₈⁻$ in 
Ref.~[\onlinecite{PhysRevB.83.165210}].
From the bases comparison I conclude that in order to translate my $Γ₈$ irrep into the notations of~Ref.[\onlinecite{PhysRevB.83.165210}] I have to
apply the following unitary transformation to my $Γ₈$ matrices:
\begin{equation}\label{GammaVIIImod}
Γ₈'=UΓ₈U^{-1},\quad U=\begin{pmatrix}0&0&-1&0\\0&0&0&1\\1&0&0&0\\0&-1&0&0\end{pmatrix}.
\end{equation}
With the matrices~$Γ₈'$ I do recover the linear basis for $Γ₈⁻$ on p.~[\onlinecite{PhysRevB.83.165210}]19:
\begin{equation}\label{linP}  \begin{split}
   [√3(iy+x)↑, -2z↑+(iy+x)↓,&\\ -2z↓+(iy-x)↑,& √3(iy-x)↓].
\end{split}\end{equation}
However, instead of~([\onlinecite{PhysRevB.83.165210}]30a) I obtain (in the notations of Ref.~[\onlinecite{PhysRevB.83.165210}]):
\begin{equation}  \begin{split}
    K_{Γ₈,Γ₈}∝\begin{pmatrix}   0&3k₊&0&√3k₋\cr -k₋&0&-√3k₊&4k₃\cr -4k₃&-√3k₊&0&-k₊\cr √3k₊&0&3k₋&0 \end{pmatrix}+\\
    +\mathrm{const}· \begin{pmatrix} 0&3k₊&6k₃&-√3k₋\cr -5k₋&0&√3k₊&2k₃\cr -2k₃&√3k₋&0&-5k₊\cr -√3k₊&-6k₃&3k₋&0\end{pmatrix},
\end{split}\end{equation}
which does not fully agree with~([\onlinecite{PhysRevB.83.165210}]30a).

Next, with the matrices~$Γ₈'$ I obtain [apart from~\eqref{linP}] two second-order polynomial bases which are incompatible with the ones  on p.~[\onlinecite{PhysRevB.83.165210}]19:
\begin{equation}\label{quadP}  \begin{split}
    [(2z²-y²-x²)↓,& √3(-y²+x²)↑,\\ √3(y²-x²)↓,& (-2z²+y²+x²)↑],
  \end{split}\end{equation}
and
\begin{equation}\label{quadPII}  \begin{split}
    [-√3(y+ix)z↑, -(y+ix)z↓+2xy↑,&\\ (y-ix)z↑+2xy↓, √3(y-ix)&z↓].
  \end{split}\end{equation}
The reason for this incompatibility might be the fact that the sixth basis on p.[\onlinecite{PhysRevB.83.165210}]19
\begin{equation}  \label{eq:3}  \begin{split}
    φ₁=-2yz↑+3iy²↓+2ixz↑-4xy↓-3ix²↓,\\ %
    φ₂=i√3(2z²↑-2iyz↓-y²↑-2xz↓-x²↑),\\ %
    φ₃=i√3(-2z²↓+2iyz↑+y²↓-2xz↑+x²↓),\\ %
    φ₄=2yz↓-3iy²↑+2ixz↓-4xy↑+3ix²↑ %
  \end{split}\end{equation}
is wrong. (All other bases on p.[\onlinecite{PhysRevB.83.165210}]19 are correct.\cite{suppMat})
First let me express the basis~\eqref{eq:3} as a linear combination of spherical harmonics\cite{LandauQM}:
\begin{equation}\label{badBasis}  \begin{split}
φ₁=i√{\frac{8π}{15}}\left[2Y_{2,1}↑+\left(Y_{2,2}+5Y_{2,-2}\right)↓\right],\\ %
φ₂=-4i√{\frac{π}5}\left(√2Y_{2,1}↓+√3Y_{2,0}↑\right),\\ %
φ₃=4i√{\frac{π}5}\left(√2Y_{2,-1}↑+√3Y_{2,0}↓\right),\\ %
φ₄=-i√{\frac{8π}{15}}\left[2Y_{2,-1}↓+\left(Y_{2,-2}+5Y_{2,2}\right)↑\right]. %
  \end{split}
\end{equation}
Next, one can check\cite{suppMat} that the system of functions~\eqref{badBasis} is (i) orthogonal\cite{Checked} and (ii) diagonalizes\cite{Checked} the
$\hat{\vec L}·\hat{\vec S}$ operator. According to the commonly accepted belief, this should mean that~\eqref{badBasis} is a basis of an irrep of the two groups:
$O⊗\{𝟙,\hat I\}$ and $T_d⊗\{𝟙,\hat I\}$. However, the direct check\cite{Checked} (see the file {\tt badBasis.txt} in the supplementary material\cite{suppMat})
shows that the system of functions~\eqref{badBasis} does not form a basis of any of the two considered groups.
In particular, the rotation around the $y$-axis by $π/2$ (which is a symmetry transformation of both groups) does not transform the system of
functions~\eqref{badBasis} into linear combinations of themselves.

I am not aware of a rigid proof of reliability of the standard $\hat{\vec L}\hat{\vec S}$-diagonalization method for finding the basis of an irrep.
Moreover, the system of functions~\eqref{badBasis} is an example which demonstrates unreliability of the commonly accepted method of obtaining the polynomial
bases of irreps of double groups.

On the contrary, the situation with the method of projection operators\cite{PeTri,DresselhausGT} (used in this article) is clear:
projection operators can never produce a wrong basis of an irrep.
\begin{widetext}
\begin{center}    \begin{table}
      \begin{tabular}{|C|C|C|C|C|C|C|C|C|C|C|C|C|C|C|C|C|}\hline
        Γ^±₁&1&1&1&1&1&1&1& 1&±1&±1&±1&±1&±1&±1&±1&±1\tabularnewline\hline
        Γ^±₂&1&1&-1&-1&1&1&-1&1&±1&±1&∓1&∓1&±1&±1&∓1&±1\tabularnewline\hline
        Γ^±₃&2&2&0&0&-1&2&0&-1&±2&±2&0&0&∓1&±2&0&∓1\tabularnewline\hline
        Γ^±₄&3&-1&-1&1&0&3&-1&0&±3&∓1&∓1&±1&0&±3&∓1&0\tabularnewline\hline
        Γ^±₅&3&-1&1&-1&0&3&1&0&±3&∓1&±1&∓1&0&±3&±1&0\tabularnewline\hline
        Γ^±₆&2&0&√2&0&1&-2&-√2&-1&±2i&0&±√2i&0&±i&∓2i&∓√2i&∓i\tabularnewline\hline
        Γ^±₇&2&0&-√2&0&1&-2&√2&-1&±2i&0&∓√2i&0&±i&∓2i&±√2i&∓i\tabularnewline\hline
        Γ^±₈&4&0&0&0&-1&-4&0&1&±4i&0&0&0&∓i&∓4i&0&±i\tabularnewline\hline
      \end{tabular}
    \caption{Characters of the  irreps of the double $O⊗\{𝟙,\hat I\}$ group in Cartan gauge.  (The inversion operator $I$ multiplies a spinor by $i$.)
The superscript $±$ in the irreps-notation means that all its polynomial bases are even/odd with respect to the transformation~$(x,y,z)→(-x,-y,-z)$.
The 3D vectors are transformed according to the irrep $Γ₅⁻$; for the spinors the appropriate irrep is $Γ₆⁺$.
   \label{tab:OhOCartan}}
  \end{table}
  \begin{table}
    \begin{tabular}{|C|C|C|C|C|C|C|C|C|C|C|C|C|C|C|C|C|}\hline
      Γ^±₁&1&1&1&1&1&1&1& 1&±1&±1&±1&±1&±1&±1&±1&±1\tabularnewline\hline
      Γ^±₂&1&1&-1&-1&1&1&-1&1&±1&±1&∓1&∓1&±1&±1&∓1&±1\tabularnewline\hline
      Γ^±₃&2&2&0&0&-1&2&0&-1&±2&±2&0&0&∓1&±2&0&∓1\tabularnewline\hline
      Γ^±₄&3&-1&-1&1&0&3&-1&0&±3&∓1&∓1&±1&0&±3&∓1&0\tabularnewline\hline
      Γ^±₅&3&-1&1&-1&0&3&1&0&3&∓1&±1&∓1&0&±3&±1&0\tabularnewline\hline
      Γ^±₆&2&0&-√2i&0&1&-2&√2i&-1&±2i&0&∓√2&0&±i&∓2i&±√2&∓i\tabularnewline\hline
      Γ^±₇&2&0&√2i&0&1&-2&-√2i&-1&±2i&0&±√2&0&±i&∓2i&∓√2&∓i\tabularnewline\hline
      Γ^±₈&4&0&0&0&-1&-4&0&1&±4i&0&0&0&∓i&∓4i&0&±i\tabularnewline\hline
    \end{tabular}
    \caption{Characters of the  irreps of the double $T_d⊗\{𝟙,\hat I\}$ group in Cartan gauge.
The 3D vectors are transformed according to the irrep $Γ₄⁻$; for the spinors the appropriate irrep is $Γ₆⁺$.\label{tab:OhTdCartan}}\end{table}
\end{center}
\end{widetext}

\section{Conclusion}\label{sec:CheckMe}
From the  application point of view, the most important result of the article is the general form of the matrix elements of the  $\vec k·\hat{\vec p}$-operator.
I expect it to be useful for studying effects in many important materials, for example, in
\begin{itemize}
\item zinc-blende semiconductors (GaAs, InAs, etc.),
\item semiconductors with inversion center (Ge, Si, etc.), and
\item cubic perovskites (SrTiO₃, LaAlO₃, etc.).
\end{itemize}
There are, however, other important (and more fundamental) results discussed in the next two paragraphs.
\paragraph{The matrices of the irreducible representations of the double cubic groups $O$, $T_d$, $O⊗\{𝟙,\hat I\}$ and $T_d⊗\{𝟙,\hat I\}$ are obtained.}
First, these matrices are important for the derivation of the bases of irreducible representations, which is useful both in analytical and in numerical
calculations. This is the only reliable method for the bases derivation I am aware of.
An alternative approach is used in Ref.[\onlinecite{PhysRevB.83.165210}] and is based on the statement (which seems to be commonly believed) that any linear combination of spherical harmonics $Y_{lm}$ with the same value of $l$ which
diagonalizes the $\hat{\vec L}\!·\!\hat{\vec S}$-operator should be a basis for some irreducible representation.
In Sec.~\ref{sec:comparison} I disprove this statement using a counter-example.
There is somewhat similar method used in Ref.~[\onlinecite{DresselhausGT}] to construct a linear basis for the $Γ₈$-irrep of the double group~$T_d$,
which assumes, however, that at least one basis function is known or can be guessed.
Having tried to guess elements of the polynomial bases which I have obtained (see Sec.~\ref{sec:bases} and Ref.~[\onlinecite{suppMat}])
I conclude that this ``guessing'' requirement is almost never fulfilled, so the method\cite{DresselhausGT} is practically useless.

\paragraph{An algorithm for obtaining general matrix structure of matrix elements between degenerate energy levels is developed.}
This algorithm is more reliable than obtaining the matrix elements from the some manually (casually) chosen set of  basis functions:
there is always a chance, that due to the oversimplified (not general enough) basis one obtains wrong matrix elements:
e.g., considering linear basis for the $Γ₈$-irrep of $T_d$ group, one might erroneously conclude that
$〈Γ₈|\vec k·\hat{\vec p}\,|Γ₈〉=0$.

This algorithm might be especially useful for analytical calculations with the $\vec k·\hat{\vec p}$ method;
for this purpose the general form of the $\vec k·\hat{\vec p}$ matrix elements is derived
(see Sec.\ref{sec:kpMatrixes} in the Appendix).

Note that there is an inconsistency between the results of this article and Ref.~[\onlinecite{WinklerBasis}]: the $8×8$ Kane model in
Table~[\onlinecite{WinklerBasis}]C.8 which has too many parameters (while only two are allowed).  There is also an inconsistency (of both my results
and Table~[\onlinecite{WinklerBasis}]C.8) with matrix elements obtained in Ref.~[\onlinecite{PhysRevB.83.165210}], but I believe that it is connected with the
invalidity of the basis\cite{PhysRevB.83.165210}, see Sec.~\ref{sec:comparison}.
\vspace{2ex}\hrule\vspace{2ex}
As a final note, let me mention somewhat misleading notation $O_h$ which is often used for both  $O⊗\{𝟙,\hat I\}$ and $T_d⊗\{𝟙,\hat I\}$ double groups.
On the first glance it makes no sense to consider the double groups $O⊗\{𝟙,\hat I\}$ and $T_d⊗\{𝟙,\hat I\}$ separately:
for example, in Pauli gauge these groups are isomorphic; moreover, any polynomial basis of some irrep
of the  double group $O⊗\{𝟙,\hat I\}$ is also a basis of some irrep
of the  double group $T_d⊗\{𝟙,\hat I\}$. However, these bases often correspond to \emph{different} irreducible representations, so it would
be incorrect to claim that two double groups $O⊗\{𝟙,\hat I\}$ and $T_d⊗\{𝟙,\hat I\}$ are identical (no matter what gauge is used).
See also the note in the end of Sec.~\ref{sec:bases:Td}.

\section{Appendix}
\subsection{Notations}\label{sec:notations}
I assume that $√{±i}=\exp[±iπ/4]$,
$(σ₁,σ₂,σ₃)$ are the usual Pauli matrices,
 $ℕ$ is the set of all positive integers, and $ℂ$ is the set of complex numbers.
By saying that ``two matrices are similar'' I mean that they are connected by some similarity transformation.
I denote $𝟙$ as an identity operator, and~$\hat I$  as the spatial inversion operator.

The irreducible representations are named according to the following rules:
\begin{itemize}
\item An irrep named $Γ₆$ (or $Γ₆⁺$ for the groups with inversion) should have a basis $[↑,↓]$.
\item All polynomial bases of ``even'' irreps (marked with the ``plus'' sign) must be invariant with respect to the transformation
$(x,y,z)→(-x,-y,-z)$.
\end{itemize}
Equations from external sources are cited, as ([{\tt citation}]NN), where NN is the equation number. For example,
([\onlinecite{PhysRevB.83.165210}]31) stands for the ``equation (31) in the article [\onlinecite{PhysRevB.83.165210}]''.

\begin{widetext}
  \subsection{General form of the $\vec k·\hat{\vec p}$ matrix elements}\label{sec:kpMatrixes}
  \subsubsection{$\vec k·\hat{\vec p}$ matrix elements for the double group $O$}\label{s:a:kp:O}
  \begin{equation}
    〈Γ₆|\vec k·\hat{\vec p}\,|Γ₆〉∝\vec k·\vec{σ},\quad    〈Γ₆|\vec k·\hat{\vec p}\,|Γ₇〉=0,\quad  〈Γ₇|\vec k·\hat{\vec p}\,|Γ₇〉∝\vec k·{\vec{σ}}^*,
  \end{equation}
  \begin{equation}      〈Γ₆|\vec k·\hat{\vec p}\,|Γ₈〉∝
k₁\begin{pmatrix}-√3&0&1&0\cr 0&-1&0&√3\end{pmatrix}+
k₂\begin{pmatrix}-√3i&0&-i&0\cr 0&-i&0&-√3i\end{pmatrix}+
k₃\begin{pmatrix}0&2&0&0\cr 0&0&2&0\end{pmatrix},
    \end{equation}
  \begin{equation}      〈Γ₇|\vec k·\hat{\vec p}\,|Γ₈〉∝
k₁\begin{pmatrix}0&√3&0&1\cr 1&0&√3&0\end{pmatrix}+
k₂\begin{pmatrix}0&-i√3&0&i\cr -i&0&i√3&0\end{pmatrix}+
k₃\begin{pmatrix}-2&0&0&0\cr 0&0&0&2\end{pmatrix},
    \end{equation}
  \begin{equation}    \begin{split}
      〈Γ₈|\vec k·\hat{\vec p}\,|Γ₈〉 ∝
k₁\begin{pmatrix}0&√3&0&-1\cr √3&0&1&0\cr 0&1&0&√3\cr -1&0&√3&0\end{pmatrix}+
k₂\begin{pmatrix}0&-√3\,i&0&-i\cr √3\,i&0&-i&0\cr 0&i&0&-√3\,i\cr i&0&√3\,i&0\end{pmatrix}+
k₃\begin{pmatrix}2&0&0&0\cr 0&2&0&0\cr 0&0&-2&0\cr 0&0&0&-2\end{pmatrix}+\\
+\mathrm{const}·\left[
k₁\begin{pmatrix}0&-√3&0&-1\cr -√3&0&-3&0\cr 0&-3&0&-√3\cr -1&0&-√3&0\end{pmatrix}+
k₂\begin{pmatrix}0&√3\,i&0&-i\cr -√3\,i&0&3\,i&0\cr 0&-3\,i&0&√3\,i\cr i&0&-√3\,i&0\end{pmatrix}+
k₃\begin{pmatrix}-4&0&0&0\cr 0&0&0&0\cr 0&0&0&0\cr 0&0&0&4\end{pmatrix}.
\right].
\end{split}  \end{equation}

\subsubsection{$\vec k·\hat{\vec p}$ matrix elements for the double group $T_d$}\label{s:a:kp:Td}
I have obtained bases and matrix elements of the double group $T_d$ in both Pauli and Cartan gauges (see Table~\ref{tab:TdCharactersCartan} for the Cartan
characters). The results are gauge-independent, as predicted by theory:\cite{Altmann,Zhelnorowich,Joshi}
  \begin{equation}
    〈Γ₆|\vec k·\hat{\vec p}\,|Γ₆〉=0,\quad     〈Γ₇|\vec k·\hat{\vec p}\,|Γ₇〉=0.
  \end{equation}
  Matrix elements (MEs) $〈Γ₆|\vec k·\hat{\vec p}\,|Γ₇〉$ are written in~\eqref{m647}.
  \begin{equation}
      〈Γ₆|\vec k·\hat{\vec p}\,|Γ₈〉∝
k₁\begin{pmatrix}1&0&√3&0\cr 0&-√3&0&-1\end{pmatrix}+
k₂\begin{pmatrix}-i&0&√3\,i&0\cr 0&√3\,i&0&-i\end{pmatrix}+
k₃\begin{pmatrix}0&0&0&2\cr 2&0&0&0\end{pmatrix},
  \end{equation}
  \begin{equation}
      〈Γ₇|\vec k·\hat{\vec p}\,|Γ₈〉∝
k₁\begin{pmatrix}0&1&0&-√3\cr -√3&0&1&0\end{pmatrix}+
k₂\begin{pmatrix}0&i&0&√3\,i\cr -√3\,i&0&-i&0\end{pmatrix}+
k₃\begin{pmatrix}0&0&-2&0\cr 0&2&0&0\end{pmatrix}.
  \end{equation}
Matrix elements $〈Γ₈|\vec k·\hat{\vec p}\,|Γ₈〉$ are written in~\eqref{m848}.

\subsubsection{$\vec k·\hat{\vec p}$ matrix elements for the double group $O⊗\{𝟙,\hat I\}$}\label{sec:OxI}
Matrix elements between both odd or both even parity states are zero, e.g.,
\begin{equation}
  〈Γ₆⁺|\vec k·\hat{\vec p}\,|Γ₆⁺〉=〈Γ₆⁻|\vec k·\hat{\vec p}\,|Γ₇⁻〉=0.
\end{equation}
The other matrix elements are
\begin{equation}
〈Γ⁺₆|\vec k·\hat{\vec p}\,|Γ⁻₇〉=0,\quad  〈Γ⁺₆|\vec k·\hat{\vec p}\,|Γ⁻₆〉∝(\vec k·\vec{σ})σ₁,\quad   〈Γ⁺₇|\vec k·\hat{\vec p}\,|Γ⁻₇〉∝(k₁σ₁+k₂σ₂-k₃σ₃)σ₂,
\end{equation}
\begin{equation}
    〈Γ⁺₆|\vec k·\hat{\vec p}\,|Γ⁻₈〉
    ∝k₁\begin{pmatrix}√3&-i&0&0\cr 0&0&0&2\end{pmatrix}
    +k₂\begin{pmatrix}0&2&0&0\cr 0&0&-i√3&-i\end{pmatrix}
    +k₃\begin{pmatrix}0&0&√3&-1\cr √3&i&0&0\end{pmatrix},
\end{equation}
\begin{equation}
    〈Γ⁺₇|\vec k·\hat{\vec p}\,|Γ⁻₈〉    ∝
k₁\begin{pmatrix}1&i√3&0&0\cr 0&0&2&0\end{pmatrix}+
k₂\begin{pmatrix}2i&0&0&0\cr 0&0&-i&i√3\end{pmatrix}+
k₃\begin{pmatrix}0&0&1&√3\cr 1&-i√3&0&0\end{pmatrix},
  \end{equation}
\begin{equation}
  \begin{split}
    〈Γ⁺₈|\vec k·\hat{\vec p}\,|Γ⁻₈〉 ∝k₁\begin{pmatrix}√3&i&0&0\cr 0&0&√3&-1\cr 0&0&i&i√3\cr -i&-√3&0&0\end{pmatrix}+ k₂\begin{pmatrix}0&-2&0&0\cr 0&0&-i√3&-i\cr 0&0&-1&√3\cr
      2&0&0&0\end{pmatrix}+
    k₃\begin{pmatrix}0&0&0&-2\cr -√3&i&0&0\cr -i&√3&0&0\cr 0&0&2i&0\end{pmatrix}+\\
    \mathrm{const}·\left[
k₁\begin{pmatrix}0&-2&0&0\cr 0&0&√3\,i&i\cr 0&0&-1&√3\cr -2&0&0&0\end{pmatrix}+
k₂\begin{pmatrix}0&2\,i&0&0\cr 0&0&-√3&-1\cr 0&0&-i&√3\,i\cr 2\,i&0&0&0\end{pmatrix}+
k₃\begin{pmatrix}0&0&-√3\,i&-i\cr 0&-2&0&0\cr -2&0&0&0\cr 0&0&1&-√3\end{pmatrix}\right].
  \end{split}\end{equation}
\subsubsection{$\vec k·\hat{\vec p}$ matrix elements for the double group $T_d⊗\{𝟙,\hat I\}$}\label{sec:TdxI}
Like in Sec.~\ref{sec:OxI}, many MEs are zero by parity, and we do not write them here. The other MEs are
\begin{equation}
  〈Γ⁺₆|\vec k·\hat{\vec p}\,|Γ⁻₆〉=0,\quad    〈Γ⁺₇|\vec k·\hat{\vec p}\,|Γ⁻₇〉=0,
\end{equation}
\begin{equation}
〈Γ⁺₆|\vec k·\hat{\vec p}\,|Γ⁻₇〉∝(\vec k·\vec{σ})σ₁\quad   〈Γ⁺₇|\vec k·\hat{\vec p}\,|Γ⁻₆〉∝(k₁σ₁+k₂σ₂-k₃σ₃)σ₂,
\end{equation}
\begin{equation}
    〈Γ⁺₆|\vec k·\hat{\vec p}\,|Γ⁻₈〉
    ∝k₁\begin{pmatrix}√3&-i&0&0\cr 0&0&0&2\end{pmatrix}
    +k₂\begin{pmatrix}0&2&0&0\cr 0&0&-i√3&-i\end{pmatrix}
    +k₃\begin{pmatrix}0&0&√3&-1\cr √3&i&0&0\end{pmatrix},
  \end{equation}
\begin{equation}
    〈Γ⁺₇|\vec k·\hat{\vec p}\,|Γ⁻₈〉
    ∝k₁\begin{pmatrix}1&i√3&0&0\cr 0&0&2&0\end{pmatrix}
    +k₂\begin{pmatrix}2i&0&0&0\cr 0&0&-i&i√3\end{pmatrix}
    +k₃\begin{pmatrix}0&0&1&√3\cr 1&-i√3&0&0\end{pmatrix},
  \end{equation}
\begin{equation}  \begin{split}
    〈Γ⁺₈|\vec k·\hat{\vec p}\,|Γ⁻₈〉   ∝
k₁\begin{pmatrix}√3&i&0&0\cr 0&0&√3&-1\cr 0&0&i&i√3\cr -i&-√3&0&0\end{pmatrix}+
k₂\begin{pmatrix}0&-2&0&0\cr 0&0&-i√3&-i\cr 0&0&-1&√3\cr 2&0&0&0\end{pmatrix}+
k₃\begin{pmatrix}0&0&0&-2\cr -√3&i&0&0\cr -i&√3&0&0\cr 0&0&2i&0\end{pmatrix}+\\
 + \mathrm{const}·\left[
k₁\begin{pmatrix}0&-2&0&0\cr 0&0&√3i&i\cr 0&0&-1&√3\cr -2&0&0&0\end{pmatrix}+
k₂\begin{pmatrix}0&2i&0&0\cr 0&0&-√3&-1\cr 0&0&-i&√3i\cr 2i&0&0&0\end{pmatrix}+
k₃\begin{pmatrix}0&0&-√3i&-i\cr 0&-2&0&0\cr -2&0&0&0\cr 0&0&1&-√3\end{pmatrix}\right].
  \end{split}\end{equation}

\subsection{The polynomial bases}\label{sec:bases}
With the information from Sec.~\ref{sec:allMatricesO} it is straightforward to get the bases for the irreps using
the standard projection operators technique (see, e.g, pp.[\onlinecite{PeTri}]82-83 or pp.[\onlinecite{DresselhausGT}]64-65).

First, let me mention some (scalar) characteristic polynomials of the irrep $Γ₁$:
\begin{equation}\tag{Γ₁}\label{GammaI}
 ψ¹₀=1,\quad  ψ¹₁=x^{2n}+y^{2n}+z^{2n},\quad ψ¹₂=y^{2n}z²+y²z^{2n}+x^{2n}z²+x^{2n}y²+x²z^{2n}+x²y^{2n},\quad ψ¹₃=(ψ²)^{2n},
\end{equation}
where $n∈ℕ$ is a positive integer\footnote{I have demonstrated the validity of~\eqref{GammaI} and~\eqref{GammaII} for~$n≤5$; for the bases of double
  groups the check has been performed mostly for~$n=1$ and~$n=2$, in some cases also for $n=3$.}, and $ψ²$ denotes an arbitrary characteristic
polynomial of the $Γ₂$-irrep, e.g.,
\begin{equation}\tag{Γ₂}\label{GammaII}
   ψ²₁=xyz,\quad ψ²₂=y²z^{2n}-x²z^{2n} -y^{2n}z²+x^{2n}z²+x²y^{2n}-x^{2n}y².
\end{equation}
Every basis below remains valid if all its components are multiplied by any expression from~\eqref{GammaI}.
The presence of such bases is considered as obvious, so they are not explicitly mentioned below.

In the main text of the article I write only few simplest polynomial bases for each irrep.
(In the supplementary material\cite{suppMat} the polynomial bases are available up to the sixth order.)

Three-dimensional vectors are transformed differently  in $O$ and $T_d$ groups.
In the $O$-case, the corresponding irrep is $Γ₅$;
in the $T_d$-case, 3D-vectors are transformed according to the  irrep  $Γ₄$.
(See the captions of the Tables~\ref{tab:OhOCartan} and~\ref{tab:OhTdCartan} for the groups with inversion centers.)
As a result, despite the fact that the two groups are isomorphic in Pauli gauge, they still have different
bases and different $\vec k·\hat{\vec p}$ matrix elements.

\subsubsection{Polynomial bases of the double group~$O$}
For $Γ₆$:
\begin{equation}\label{OGammaVI}  \begin{split}
    [↑,↓],\quad [-z^{2n-1}↑-iy^{2n-1}↓-x^{2n-1}↓, z^{2n-1}↓+iy^{2n-1}↑-x^{2n-1}↑],\quad n∈ℕ,\\
    ψ²[yz↓+ixz↓+xy↑, yz↑-ixz↑-xy↓],
\end{split}  \end{equation}
where $ψ²$ denotes an arbitrary characteristic polynomial of  the $Γ₂$-irrep, taken, e.g.,  from~\eqref{GammaII}.
For $Γ₇$:
\begin{equation}\label{OGammaVII}  \begin{split}
    ψ²&[↓,-↑],\quad    ψ²[z^{2n-1}↓+iy^{2n-1}↑-x^{2n-1}↑, z^{2n-1}↑+iy^{2n-1}↓+x^{2n-1}↓],\\
    &[-(yz)^{2n-1}↑+i(xz)^{2n-1}↑+(xy)^{2n-1}↓, (yz)^{2n-1}↓+i(xz)^{2n-1}↓+(xy)^{2n-1}↑].
  \end{split}\end{equation}
For $Γ₈$:
  \begin{equation}\label{OGammaVIII}
    \begin{split}
[-√3(x^{2n-1}+iy^{2n-1})↑, &2z^{2n-1}↑-iy^{2n-1}↓-x^{2n-1}↓,\\ &2z^{2n-1}↓-iy^{2n-1}↑+x^{2n-1}↑, √3(x^{2n-1}-iy^{2n-1})↓],\\
ψ²[2z↓+x↑-iy↑,& √3(-x+iy)↓, √3(x+iy)↑, 2z↑-x↓-iy↓],\\
ψ²[√3z(y+ix)↑,& yz↓+ixz↓-2xy↑, -yz↑+ixz↑-2xy↓, √3z(-y+ix)↓],\\
[√3(-y^{2n}+x^{2n})↓,& (-2z^{2n}+y^{2n}+x^{2n})↑, (2z^{2n}-y^{2n}-x^{2n})↓, √3(y^{2n}-x^{2n})↑].
    \end{split}
\end{equation}

\subsubsection{Polynomial bases of the double group~$T_d$}\label{sec:bases:Td}
For $Γ₆$:
\begin{equation}\label{TdGammaVI}  \begin{split}
[↑,↓],\quad &ψ²[↑,↓],\quad ψ²[z³↑+y³↑+x³↑, z³↓+y³↓+x³↓],\\
[yz^{2n}↓&+iy^{2n}z↑+ixz^{2n}↓-ixy^{2n}↓-ix^{2n}z↑-x^{2n}y↓,\\ &-yz^{2n}↑-iy^{2n}z↓+ixz^{2n}↑-ixy^{2n}↑+ix^{2n}z↓+x^{2n}y↑].
  \end{split}\end{equation}
For $Γ₇$:
\begin{equation}\label{TdGammaVII}
  \begin{split}
&[z^{2n-1}↓+iy^{2n-1}↑-x^{2n-1}↑, z^{2n-1}↑+iy^{2n-1}↓+x^{2n-1}↓],\quad [-yz↑+ixz↑+xy↓, yz↓+ixz↓+xy↑],\\
ψ²&[z^{2n-1}↓+iy^{2n-1}↑-x^{2n-1}↑, z^{2n-1}↑+iy^{2n-1}↓+x^{2n-1}↓],\quad ψ²[-yz↑+ixz↑+xy↓, yz↓+ixz↓+xy↑].
  \end{split} \end{equation}
For $Γ₈$:
\begin{equation} \label{TdGammaVIII}   \begin{split}
  [-yz↑+ixz↑-2xy↓, √3z(y-x)↓,& -√3z(y-ix)↑, yz↓+ixz↓-2xy↑],\\
ψ²[-yz↑+ixz↑-2xy↓, √3z(y-x)↓,& -√3z(y-ix)↑, yz↓+ixz↓-2xy↑],\\
  [2z^{2n-1}↓-iy^{2n-1}↑+x^{2n-1}↑,& √3(iy^{2n-1}-x^{2n-1})↓,\\ &√3(iy^{2n-1}+x^{2n-1})↑, 2z^{2n-1}↑-iy^{2n-1}↓-x^{2n-1}↓],\\
  [√3(-y^{2n}+x^{2n})↓,(-2z^{2n}+y^{2n}&+x^{2n})↑,(2z^{2n}-y^{2n}-x^{2n})↓, √3(y^{2n}-x^{2n})↑],\\
[2yz^{2n}↑+iy^{2n}z↓+2ixz^{2n}↑-ixy^{2n}↑+ix^{2n}z↓&-x^{2n}y↑,√3(-iy^{2n}z↑-ixy^{2n}↓+ix^{2n}z↑-x^{2n}y↓),\\ √3(-iy^{2n}z↓+ixy^{2n}↑+ix^{2n}z↓-x^{2n}y↑),&
2yz^{2n}↓+iy^{2n}z↑-2ixz^{2n}↓+ixy^{2n}↓+ix^{2n}z↑-x^{2n}y↓].
 \end{split}\end{equation}
Note that the two bases~\eqref{OGammaVIII} and~\eqref{TdGammaVIII} are not equivalent, because the order of the functions in an irrep-basis \emph{matters}:
e.g., one could interchange elements in the first basis set in~\eqref{TdGammaVIII} in such a way that it becomes equal to the first basis set in~\eqref{OGammaVIII};
however, such an exchange would make second basis sets in~\eqref{TdGammaVIII} and in~\eqref{OGammaVIII} different.
Generally, a wave function is given by a mixture of several basis sets.

\subsubsection{Polynomial bases of the double group~$O⊗\{𝟙,\hat I\}$}
In case of $O⊗\{𝟙,\hat I\}$ the vectors/spinors are transformed according to the $Γ₅⁻$/$Γ₆⁺$ irreps.
The bases for $Γ₆^+$ are:
\begin{equation}\label{pbA}  \begin{split}
    [↑,↓],\quad     [&yz^{2n-1}↓-y^{2n-1}z↓-ixz^{2n-1}↓+xy^{2n-1}↑+ix^{2n-1}z↓-x^{2n-1}y↑,\\
& yz^{2n-1}↑-y^{2n-1}z↑+ixz^{2n-1}↑-xy^{2n-1}↓-ix^{2n-1}z↑+x^{2n-1}y↓].
\end{split}\end{equation}
For $Γ₆^-$:
\begin{equation}\begin{split}
[z^{2n-1}↓+iy^{2n-1}↑-x^{2n-1}↑,& -z^{2n-1}↑-iy^{2n-1}↓-x^{2n-1}↓],\\
[-yz^{2n}↑-ixz^{2n}↑-ixy^{2n}↑&+iy^{2n}z↓+ix^{2n}z↓-x^{2n}y↑,\\ &yz^{2n}↓-iy^{2n}z↑-ixz^{2n}↓-ixy^{2n}↓-ix^{2n}z↑+x^{2n}y↓],\\
ψ²[yz↑-ixz↑-xy↓,& yz↓+ixz↓+xy↑].
\end{split} \label{defFz} \end{equation}
For $Γ₇^+$:
  \begin{equation}  \begin{split}
[(yz)^{2n-1}↓+i(xz)^{2n-1}↓+(xy)^{2n-1}↑,& -(yz)^{2n-1}↑+i(xz)^{2n-1}↑+(xy)^{2n-1}↓],\\
[yz^{2n-1}↓+y^{2n-1}z↓+ixz^{2n-1}↓&+xy^{2n-1}↑+ix^{2n-1}z↓+x^{2n-1}y↑,\\ -yz^{2n-1}↑-y^{2n-1}z↑&+ixz^{2n-1}↑+xy^{2n-1}↓+ix^{2n-1}z↑+x^{2n-1}y↓],\\
ψ²[z^{2n-1}↑+iy^{2n-1}↓+x^{2n-1}↓,& z^{2n-1}↓+iy^{2n-1}↑-x^{2n-1}↑].
    \end{split}\end{equation}
For $Γ₇^-$:
\begin{equation}  \begin{split}
ψ² [↓,↑],\quad 
[&-yz^{2n}↑-iy^{2n}z↓+ixz^{2n}↑-ixy^{2n}↑+ix^{2n}z↓+x^{2n}y↑,\\ &yz^{2n}↓+iy^{2n}z↑+ixz^{2n}↓-ixy^{2n}↓-ix^{2n}z↑-x^{2n}y↓].
  \end{split}\end{equation}
For $Γ₈^+$:
\begin{equation}  \begin{split}
[-z^{2n}↑+2y^{2n}↑-x^{2n}↑,& z^{2n}↓-2y^{2n}↓+x^{2n}↓,i√3(-z^{2n}+x^{2n})↓, i√3(z^{2n}↑-x^{2n})↑], \\
ψ²[√3(z^{2n-1}↑-x^{2n-1}↓),& √3(z^{2n-1}↓+√3x^{2n-1}↑),\\ &iz^{2n-1}↓+2y^{2n-1}↑-ix^{2n-1}↑, iz^{2n-1}↑+2y^{2n-1}↓+ix^{2n-1}↓], \\
[√3(yz↓-xy↑),& -√3(yz↑+xy↓), iyz↑-2xz↑-ixy↓, -iyz↓-2xz↓-ixy↑].
\end{split}\end{equation}
For $Γ₈^-$:
\begin{equation}  \begin{split}
[√3(z^{2n-1}↓+x^{2n-1}↑),& iz^{2n-1}↓+2y^{2n-1}↑-ix^{2n-1}↑,\\ & √3(z^{2n-1}↑-iy^{2n-1}↓), -z^{2n-1}↑-iy^{2n-1}↓+2x^{2n-1}↓],\\
ψ²[z²↓-2y²↓+x²↓,&i√3(-z²↓+x²)↓, z²↑+y²↑-2x²↑,√3(z²-y²)↑],\\
ψ²[√3(yz↑+xy↓),& -iyz↑+2xz↑+ixy↓, √3(-ixz↓+xy↑), 2yz↓-ixz↓-xy↑],\\
[√3(-yz^{2n}↑-ixz^{2n}↑&-ix^{2n}z↓+x^{2n}y↑), -iyz^{2n}↑+2y^{2n}z↓+xz^{2n}↑-2xy^{2n}↑-x^{2n}z↓-ix^{2n}y↑,\\
 √3(-yz^{2n}↓-iy^{2n}z↑&+ixz^{2n}↓-ixy^{2n}↓), yz^{2n}↓-iy^{2n}z↑-ixz^{2n}↓-ixy^{2n}↓+2ix^{2n}z↑-2x^{2n}y↓].
\end{split}\end{equation}

\subsubsection{Polynomial bases of the double group~$T_d⊗\{𝟙,\hat I\}$}
In case of $T_d⊗\{𝟙,\hat I\}$ the vectors/spinors are transformed according to the $Γ₄⁻$/$Γ₇⁺$ irreps. The bases for $Γ₆^+$ are:
\begin{equation}
  \begin{split}
    [↑,↓],\quad [&yz^{2n-1}↓-y^{2n-1}z↓-ixz^{2n-1}↓+xy^{2n-1}↑+ix^{2n-1}z↓-x^{2n-1}y↑,\\ & yz^{2n-1}↑-y^{2n-1}z↑+ixz^{2n-1}↑-xy^{2n-1}↓-ix^{2n-1}z↑+x^{2n-1}y↓].
  \end{split}
\end{equation}
For $Γ₆^-$:
\begin{equation}\begin{split}
ψ²[↓,↑],\quad [&-yz^{2n}↑-iy^{2n}z↓+ixz^{2n}↑-ixy^{2n}↑+ix^{2n}z↓+x^{2n}y↑,\\ & yz^{2n}↓+iy^{2n}z↑+ixz^{2n}↓-ixy^{2n}↓-ix^{2n}z↑-x^{2n}y↓].
\end{split} \end{equation}
For $Γ₇^+$:
  \begin{equation}    \begin{split}
&[yz↓+ixz↓+xy↑, -yz↑+ixz↑+xy↓],\\  &ψ²[z^{2n-1}↑+iy^{2n-1}↓+x^{2n-1}↓, z^{2n-1}↓+iy^{2n-1}↑-x^{2n-1}↑].
    \end{split}\end{equation}
For $Γ₇^-$:
\begin{equation}  \begin{split}
[z^{2n-1}↓+iy^{2n-1}↑-x^{2n-1}↑,& -z^{2n-1}↑-iy^{2n-1}↓-x^{2n-1}↓],\\
[-yz^{2n}↑+iy^{2n}z↓-ixz^{2n}↑&-ixy^{2n}↑+ix^{2n}z↓-x^{2n}y↑,\\ yz^{2n}↓&-iy^{2n}z↑-ixz^{2n}↓-ixy^{2n}↓-ix^{2n}z↑+x^{2n}y↓],\\
ψ²[yz↑-ixz↑-xy↓,& yz↓+ixz↓+xy↑].
  \end{split}\end{equation}
For $Γ₈^+$:
\begin{equation}  \begin{split}
[-z^{2n}↑+2y^{2n}↑-x^{2n}↑, z^{2n}↓&-2y^{2n}↓+x^{2n}↓, i√3(-z^{2n}+x^{2n})↓, i√3(z^{2n}-x^{2n})↑],\\
ψ²[√3(z^{2n-1}↑-x^{2n-1}↓),& √3(z^{2n-1}↓+x^{2n-1}↑),\\ &iz^{2n-1}↓+2y^{2n-1}↑-ix^{2n-1}↑, iz^{2n-1}↑+2y^{2n-1}↓+ix^{2n-1}↓],\\
[√3\{(yz)^{2n-1}↓-(xy)^{2n-1}↑\},& -√3\{(yz)^{2n-1}↑+(xy)^{2n-1}↓\},\\ i(yz)^{2n-1}↑&-2(xz)^{2n-1}↑-i(xy)^{2n-1}↓, -i(yz)^{2n-1}↓-2(xz)^{2n-1}↓-i(xy)^{2n-1}↑].
\end{split}\end{equation}
For $Γ₈^-$:
\begin{equation}  \begin{split}
[√3(z^{2n-1}↓+x^{2n-1}↑),& iz^{2n-1}↓+2y^{2n-1}↑-ix^{2n-1}↑,\\ √3&(z^{2n-1}↑-√3iy^{2n-1}↓), -z^{2n-1}↑-iy^{2n-1}↓+2x^{2n-1}↓],\\
ψ²[z²↓-2y²↓+x²↓,& i√3(x²-z²)↓, z²↑+y²↑-2x²↑, √3(z²-y²)↑],\\
ψ²[√3(yz↑+xy↓),& -iyz↑+2xz↑+ixy↓, √3(-ixz↓+xy↑), 2yz↓-ixz↓-xy↑].
\end{split}\end{equation}
\end{widetext}
\Russian
\bibliography{local}
\end{document}

%% file: local.tex
\usepackage[utf8x]{inputenc}\SetUnicodeOption{mathletters} \SetUnicodeOption{autogenerated}
\PreloadUnicodePage{4}
\DeclareUnicodeCharacter{9786}{\Smiley}
\DeclareUnicodeCharacter{9785}{\frownie}
\DeclareUnicodeCharacter{183}{\cdot}
\DeclareUnicodeCharacter{9001}{\langle}
\DeclareUnicodeCharacter{9002}{\rangle}
\DeclareUnicodeCharacter{8469}{\mathbb N}
\DeclareUnicodeCharacter{8450}{\mathbb C}
\DeclareUnicodeCharacter{8730}{\sqrt}
\DeclareUnicodeCharacter{8314}{^+}
\DeclareUnicodeCharacter{8315}{^-}
\DeclareUnicodeCharacter{9829}{\heartsuit}
\newcommand{\ud}{\mathrm{d}}

\usepackage{array,graphicx,amssymb,amsmath,amsbsy,subfigure,dsfont,ifthen,ifpdf,color,subscript,marvosym,bbm} 

\newcommand{\insfig}[2]{\begin{minipage}[c]{#1}\includegraphics[width=#1]{#2}\end{minipage}}
\usepackage[russian,english]{babel}
\usepackage[unicode]{hyperref}
\hypersetup{pdfstartview=FitH,
pdftitle={Matrices and characteristic polynomials for double crystallographic point groups},%
pdfsubject={derivation of the matrices for double groups},pdfauthor={Oleg Chalaev},%
pdfkeywords={double groups,spin,crystals}}%

\newcounter{numberOfGroupElement} 

\newcounter{classNumber} 

%% file: main.bbl
\begin{thebibliography}{10}
\bibitem{Checked}%
  \BibitemOpen
  \bibinfo {note} {This statement has been directly checked on computer.}%
  \bibAnnoteFile{Stop}{Checked}%
\end{thebibliography}
